\documentclass[fleqn,usenatbib]{mnras}

\usepackage{graphicx}	
\usepackage{multicol}   
\usepackage{xspace}

\newcommand{\Jybeam}{Jy\,beam$^{-1}$\xspace}
\newcommand{\mJybeam}{mJy\,beam$^{-1}$\xspace}

\newcommand{\sigmav}{$\langle\sigma v\rangle_\mathrm{\chi}$\xspace}

\usepackage[T1]{fontenc}
\usepackage{ae,aecompl}

\usepackage{newtxtext,newtxmath}

\title[Low-Frequency Search for Dark Matter Signals in dSphs]{Searching for Dark Matter Signals from Local Dwarf Spheroidal Galaxies at Low Radio Frequencies in the GLEAM Survey}

\date{Last updated 2019 Septemeber 04}

\pubyear{2019}

\newcommand{\affil}[1]{$^{\rm #1}$}
\newcounter{inst}
\newcommand{\inst}[1]{\noindent%
   \refstepcounter{inst}\affil{\arabic{inst}\label{#1}}     
}

\author[R. H. W. Cook et al. 2019]
{R. H. W. Cook\affil{\ref{ICRAR},\ref{CIRA},\ref{ASTRO3D}}, 
N. Seymour\affil{\ref{CIRA}},
K. Spekkens\affil{\ref{RMC}},
N. Hurley-Walker\affil{\ref{CIRA}},
P. J. Hancock\affil{\ref{CIRA}},\newauthor
M. E. Bell\affil{\ref{UTS}},
J. R. Callingham\affil{\ref{ASTRON}},
B.-Q. For\affil{\ref{ICRAR},\ref{ASTRO3D}},
T. M. O. Franzen\affil{\ref{CIRA},\ref{ASTRON}},
B. M. Gaensler\affil{\ref{DIAA},\ref{ASTRO3D}},\newauthor
L. Hindson,
C. A. Jackson\affil{\ref{CIRA},\ref{ASTRON}},
M. Johnston-Hollitt\affil{\ref{CIRA}},
A. D. Kapi\'{n}ska\affil{\ref{ICRAR}},
J. Morgan\affil{\ref{CIRA}},\newauthor
A. R. Offringa\affil{\ref{ASTRON}},
P. Procopio\affil{\ref{UMel}},
L. Staveley-Smith\affil{\ref{ICRAR},\ref{ASTRO3D}},
R. B. Wayth\affil{\ref{CIRA},\ref{ASTRO3D}},
C. Wu\affil{\ref{ICRAR}},
Q. Zheng\affil{\ref{SAO}}\vspace{0.25cm}\\
{\small \inst{ICRAR}\,International Centre for Radio Astronomy Research (ICRAR), University of Western Australia, Crawley, WA 6009, Australia}\\
{\small \inst{CIRA}\,International Centre for Radio Astronomy Research (ICRAR), Curtin University, Bentley, WA 6102, Australia}\\
{\small \inst{ASTRO3D}\,ARC Centre of Excellence for All Sky Astrophysics in 3 Dimensions (ASTRO 3D)}\\
{\small \inst{RMC}\,Department of Physics and Space Science, Royal Military College of Canada}\\
{\small \inst{UTS}\,University of Technology Sydney, 15 Broadway, Ultimo NSW 2007, Australia}\\
{\small \inst{ASTRON}\,ASTRON, Netherlands Institute for Radio Astronomy, Oude Hoogeveensedijk 4, 7991 PD, Dwingeloo, The Netherlands}\\
{\small \inst{DIAA}\,Dunlap Institute for Astronomy \& Astrophysics, University of Toronto, 50 St George St, Toronto, ON, M5S 3H4, Canada}\\
{\small \inst{UMel}\,School of Physics, The University of Melbourne, Parkville, VIC 3010, Australia}\\
{\small \inst{SAO}\,Shanghai Astronomical Observatory, 80 Nandan Road, Shanghai 200030, China}\\
}

\begin{document}
\label{firstpage}
\pagerange{\pageref{firstpage}--\pageref{lastpage}}
\maketitle

\begin{abstract}
The search for emission from weakly interacting massive particle (WIMP) dark matter annihilation and decay has become a multi-pronged area of research not only targeting a diverse selection of astrophysical objects, but also taking advantage of the entire electromagnetic spectrum. The decay of WIMP particles into standard model particles has been suggested as a possible channel for synchrotron emission to be detected at low radio frequencies. Here, we present the stacking analysis of a sample of 33 dwarf spheroidal (dSph) galaxies with low-frequency (72\,--\,231\,MHz) radio images from the GaLactic and Extragalactic All-sky Murchison Widefield Array (GLEAM) survey. We produce radial surface brightness profiles of images centred upon each dSph galaxy with background radio sources masked. We remove ten fields from the stacking due to contamination from either poorly subtracted, bright radio sources or strong background gradients across the field. The remaining 23 dSph galaxies are stacked in an attempt to obtain a statistical detection of any WIMP-induced synchrotron emission in these systems. We find that the stacked radial brightness profile does not exhibit a statistically significant detection above the 95\% confidence level of $\sim 1.5$\,\mJybeam. This novel technique shows the potential of using low-frequency radio images to constrain fundamental properties of particle dark matter.
\end{abstract}



\section{Introduction}
\label{sec:introduction}
Cosmological measurements from the Wilkinson Microwave Anisotropy Probe (WMAP) have shown that the fraction of non-baryonic dark matter relative to that of the critical density of the Universe is $\Omega_\mathrm{DM}\rm{h}^2\simeq0.11$ \citep{Komatsu2011}, implying a weak interaction for such particles with a mass of order GeV. Given these constraints, the weakly interacting massive particle (WIMP) is seen as an attractive dark matter candidate (see \citealt{Jungman1996, Bergstrom2000, Bertone2005, Feng2010} for extensive reviews). The WIMP interacts both gravitationally as well as via the weak force, resulting in the thermalisation with standard model particles in the early Universe, freezing out in number as the Universe expanded. For this reason, the WIMP abundance in the current epoch is inversely proportional to a characteristic annihilation cross-section $\langle\sigma v\rangle_{\chi} \approx 3\times10^{-26}\,\rm{cm}^{3}\,\rm{s}^{-1}$ for a thermal relic; a requisite for reproducing the current dark matter density in the Universe \citep{Porter2011}.\\

The search for dark matter annihilation and decay signals has quickly evolved into a multi-frequency pursuit with radio and X-ray observations becoming increasingly complementary to previous $\gamma$-ray studies. It is thought to be possible to detect WIMPs by means of targeting non-gravitational interactions of dark matter with standard model particles. In an astrophysical context, WIMP searches revolve around the indirect detection of standard model particles --- particularly electrons, photons, and neutrinos --- that are a result of WIMP annihilations and/or decays \citep{Abdo2009}. $\gamma$-ray searches for dark matter annihilation signals in a range of astrophysical objects have proven to be useful in constraining upper limits to \sigmav. In particular, these searches target objects which are observed to be dark matter dominated via studies of their kinematics. Examples of such objects include: the Galactic centre (e.g., \citealt{Abazajian2012}; \citealt{Abdo2010a}, \citealt{Ackermann2012a}; \citealt{Dobler2010}), diffuse Galactic and extragalactic backgrounds (e.g., \citealt{Abdo2010c}; \citealt*{Papucci2010}; \citealt{Baxter2011}); as well as galaxy clusters (e.g. \citealt{Ackermann2010}; \citealt*{Ando2012}; \citealt{Han2012}; \citealt{Abramowski2014}). However, deviations from the predicted signals are observed in such systems and measurement uncertainties weaken the constraints of these studies (e.g. \citealt{Strigari2013}).

\citet{Borriello2009} attempted to measure secondary radio emission from WIMP annihilation products in the galactic halo and its substructure. They found that their constraints on the annihilation cross-section of $\langle\sigma v\rangle \sim 10^{-24}\,\rm{cm}^{3}\,\rm{s}^{-1}$ and WIMP mass of order $m_\mathrm{\chi} = 100\,GeV$, were limited by the diffuse nature of the electron population which lessened the radio signal. On a larger scale, \citet{Zhang2008} calculate the intensity and power spectrum of the cosmological background of synchrotron emission. They were able to probe dark matter masses $\gtrsim 100\,\rm{GeV}$ and annihilation cross-sections of as low as $\langle\sigma v\rangle \sim 3\times10^{-26}\,\rm{cm}^{3}\,\rm{s}^{-1}$ under modest assumptions of the dark matter clumping.

Additionally, dwarf spheroidal (dSph) galaxies of the Local Group have been recognised as appealing candidates for indirect dark matter searches (e.g., \citealt{Abdo2010b}; \citealt{Ackermann2011}; \citealt{Acciari2010}) due, in part, to their inferred dark matter abundance. Moreover, recent studies of the star formation rates in these systems, observed with the Fermi Large Area Telescope (LAT), have shown --- via extrapolation of larger, more luminous systems --- that the $\gamma$-ray flux density due to these intrinsic processes is negligible (\citealt{Ackermann2012b}; but see \citealt{Webster2014}). It is indeed possible that the nature of such signals are unrelated to their underlying dark matter, but instead due to other astrophysical non-thermal or thermal emission. Such processes should scale with star formation rate and given that there is negligible ongoing star formation observed in these systems (e.g. \citealt{Hernandez2000, Lee2009}), this would imply that any significant $\gamma$-ray signal detected is likely to be associated with dark matter particle annihilations within the halo.

The radiation emitted from the annihilation of dark matter particles in the halos of dwarf spheroidal galaxies spans the majority of the electromagnetic spectrum (see \citealt{Profumo2010} for a review). The annihilation of WIMP pairs produce $\gamma$-rays as a result of photon decay of neutral pions and internal bremsstrahlung from charged particle final states in the high-energy regime of the electromagnetic spectrum \citep{Abdo2009}. This study focuses on observations at low radio frequencies (i.e. $\lesssim$\,GHz) assuming the annihilation of dark matter particles to leptonic states (e.g. $e^{+}e^{-}$, $\mu^{+}\mu^{-}$, and $\tau^{+}\tau^{-}$) as described in \citet{Natarajan2013, Regis2015}. When observed at these frequencies, secondary emission is predicted from this population of non-thermal, energetic electrons and positrons \citep{Fornengo2012}. These electrons and positrons lose energy through various mechanisms, namely, inverse Compton scattering off starlight and cosmic microwave background (CMB) photons, bremsstrahlung in the presence of an ionized gas, and synchrotron emission in the presence of magnetic fields. The latter is common to WIMP dark matter models and indeed any dark matter candidate acting at the weak-scale. The intensity of radiation will be dependent upon the energy distribution of the resulting electrons and positrons which is itself dependent on the particular annihilation channel.

\citet{Colafrancesco2007} have produced models of the dark matter density profile for the specific case of the Draco dwarf spheroidal galaxy based upon mass models. They proceed to investigate the expected yield of electron and positron pairs produced through WIMP annihilations from which they model the spatially extended, synchrotron emission. These models predict that WIMP dark matter annihilations within such systems will produce a smooth, degree-scale intensity distribution. \citet{Spekkens2013} and \citet{Natarajan2013} endeavoured to constrain this WIMP-induced synchrotron emission from deep radio observations of four Local Group dwarf spheroidal galaxies, namely, Draco; Ursa Major II; Coma Berenices; Willman 1. They used 1.4\,GHz Green Bank Telescope (GBT) observations in conjunction with the NRAO VLA Sky Survey (NVSS) catalogue to produce source-subtracted maps with sensitivities of $\sigma_\mathrm{RMS} \lesssim 7\,$\mJybeam. They find no appreciable emission in the Ursa Major II and Willman 1 fields and concluded that the residual maps of Draco and Coma Berenices were likely contaminated by Galactic foreground emission. The angular scale of this extrinsic emission varies on a similar scale to that of the underlying synchrotron emission expected from the dwarf spheroidal galaxy. As such, they found that it was not possible to disentangle the two extended sources of emission. They provide upper limits on the WIMP annihilation cross-section of $\langle\sigma v\rangle_\mathrm{\chi} \lesssim 10^{-25}\,\rm{cm}^{3}\,s^{-1}$ from the Ursa Major II and Willman 1 observations.

\citet{Regis2015} performed deep mosaic radio observations of six local dSphs with Australia Telescope Compact Array (ATCA) in the 1.1\,--\,3.1\,GHz frequency band, achieving an RMS sensitivity better than 0.05\,\mJybeam\;in each of their fields. They find no evidence for extended emission within radii of a few arcminutes. In addition, they derive bounds on the WIMP annihilation/decay rate as a function of the mass for different annihilation/decay channels \citep{Regis2014} which are comparable to the best limits constrained with $\gamma$-ray observations. The use of multiple array detectors yielding synthesised beams of order arcminutes has advantages over single dish observations in that one can achieve more sensitive images, efficient subtraction of background radio sources, as well as being less contaminated by Galactic foreground emission.\\

In this paper we present the stacking analysis of 33 dwarf spheroidal galaxies using low radio frequency (72\,--\,231 MHz) images from the GaLactic and Extragalactic All-sky Murchison Widefield Array (GLEAM) survey. The GLEAM survey provides a balance between relatively high resolutions and low image noise which should allow us to better probe the faint, extended synchrotron emission predicted from the annihilation of dark matter WIMP particles.\\
The paper is formatted as follows: \S\ref{sec:data} outlines the data and dSph galaxy sample used in our analysis; \S\ref{sec:method} describes the image processing steps taken to produce radial brightness profiles for each of the dSph galaxies; we present the results of stacking these brightness profiles in \S\ref{sec:results} and discuss the implications and potential of this technique for future low-frequency radio continuum surveys in \S\ref{sec:discussion}.

\section{DATA}
\label{sec:data}

\begin{table*}
\centering
\caption{Basic optical properties of the selected dwarf spheroidal galaxies visible in the GLEAM sky ($\delta < +30^{\circ}$). dSphs with an * contain very bright, extended sources that have been poorly subtracted. Those marked with a $^{\dagger}$ have large-scale background gradients across their cutouts. Those with a $^\ddagger$ are within $\sim3^{\circ}$ of another dSph; such pairs naturally bear the same constellation name. This data has been sourced from \citet{McConnachie2012}, excluding Crater 2: a newly-discovered Local Group dwarf galaxy \citep{Torrealba2016}. The RMS column gives the average root mean square noise in GLEAM 200\,MHz images for each dSph galaxy and the final column is a measure of the mean background level in the cutout.}
\vspace{0.15cm}
\begin{tabular}{lllccc}
\hline
Galaxy Name & RA (J2000) & DEC (J2000) & $r_{H}$  & RMS  & Mean Background Level \\
 & & & (arcmin) & (mJy\,beam$^{-1}$) & (mJy\,beam$^{-1}$) \\
\hline\hline
  Bo$\ddot{\rm{o}}$tes I$^{\dagger\ddagger}$ & 14$^{h}$00$^{m}$06$^{s}$.0 & +14$^{\circ}$30$'$00$''$ & $12.6^{+1.0}_{-1.0}$ & 28.0 & -7.9\\
  Bo$\ddot{\rm{o}}$tes II$^{\dagger\ddagger}$ & 13$^{h}$58$^{m}$00$^{s}$.0 & +12$^{\circ}$51$'$00$''$ & $4.2^{+1.4}_{-1.4}$ & 25.0 & -12.0\\
  Carina & 06$^{h}$41$^{m}$36$^{s}$.7 & $-$50$^{\circ}$57$'$58$''$ & $8.2^{+1.2}_{-1.2}$ & 7.0 & -5.7\\
  Cetus II & 01$^{h}$17$^{m}$52$^{s}$.8 & $-$17$^{\circ}$25$'$12$''$ & $1.9^{+1.0}_{-0.5}$ & 7.0 & -3.2\\
  Columba I & 05$^{h}$31$^{m}$26$^{s}$.4 & $-$28$^{\circ}$01$'$48$''$ & $1.9^{+0.5}_{-0.4}$ & 5.0 & -0.7\\
  Coma Berenices$^{\dagger}$ & 12$^{h}$26$^{m}$59$^{s}$.0 & +23$^{\circ}$54$'$15$''$ & $6.0^{+0.6}_{-0.6}$ & 55.0 & -2.6\\
  Crater 2 & 11$^{h}$49$^{m}$14$^{s}$.4 & $-$18$^{\circ}$24$'$47$''$ & $31.2^{+2.5}_{-2.5}$ & 6.0 & -5.0\\
  Eridanus 3 & 02$^{h}$22$^{m}$45$^{s}$.5 & $-$52$^{\circ}$17$'$01$''$ & $0.54^{+0.5}_{-0.1}$ & 6.0 & -5.4\\
  Fornax & 02$^{h}$39$^{m}$59$^{s}$.3 & $-$34$^{\circ}$26$'$57$''$ & $16.6^{+1.2}_{-1.2}$ & 6.0 & -1.6\\
  Grus 1 & 22$^{h}$56$^{m}$42$^{s}$.4 & $-$50$^{\circ}$09$'$48$''$ & $1.77^{+0.85}_{-0.39}$ & 8.0 & -3.9\\
  Grus II & 22$^{h}$04$^{m}$04$^{s}$.8 & $-$46$^{\circ}$26$'$24$''$ & $6.0^{+0.9}_{-0.5}$ & 8.0 & -0.9\\
  Hercules$^{\dagger}$ & 16$^{h}$31$^{m}$02$^{s}$.0 & +12$^{\circ}$47$'$30$''$ & $8.6^{+1.8}_{-1.1}$ & 31.0 & -7.8\\
  Horologium 1 & 02$^{h}$55$^{m}$31$^{s}$.7 & $-$54$^{\circ}$07$'$08$''$ & $1.31^{+0.19}_{-0.14}$ & 6.0 & -2.3\\
  Horologium II & 03$^{h}$16$^{m}$32$^{s}$.1 & $-$50$^{\circ}$01$'$05$''$ & $2.09^{+0.44}_{-0.41}$ & 6.0 & -3.6\\
  Hydra II & 12$^{h}$21$^{m}$42$^{s}$.1 & $-$31$^{\circ}$59$'$07$''$ & $1.7^{+0.3}_{-0.2}$ & 7.0 & -6.7\\
  Indus I & 21$^{h}$08$^{m}$49$^{s}$.1 & $-$51$^{\circ}$09$'$56$''$ & $1.26^{+0.45}_{-0.27}$ & 9.0 & -2.0\\
  Kim 2 & 20$^{h}$38$^{m}$52$^{s}$.8 & $-$46$^{\circ}$09$'$36$''$ & $2.9^{+1.1}_{-1.0}$ & 10.0 & -10.4\\
  Leo I & 10$^{h}$08$^{m}$28$^{s}$.1 & +12$^{\circ}$18$'$23$''$ & $3.4^{+0.3}_{-0.3}$ & 14.0 & -1.2\\
  Leo II$^{\dagger}$ & 11$^{h}$13$^{m}$28$^{s}$.8 & +22$^{\circ}$09$'$06$''$ & $2.6^{+0.6}_{-0.6}$ & 18.0 & 3.1\\
  Leo IV$^{\dagger\ddagger}$ & 11$^{h}$32$^{m}$57$^{s}$.0 & $-$00$^{\circ}$32$'$00$''$ & $4.6^{+0.8}_{-0.8}$ & 9.0 & 1.1\\
  Leo V$^{\dagger\ddagger}$ & 11$^{h}$31$^{m}$09$^{s}$.6 & +02$^{\circ}$13$'$12$''$ & $2.6^{+0.6}_{-0.6}$ & 9.0 & -3.4\\
  Phoenix 2 & 23$^{h}$39$^{m}$59$^{s}$.4 & $-$54$^{\circ}$24$'$22$''$ & $1.09^{+0.26}_{-0.16}$ & 8.0 & -1.4\\
  Pictoris 1 & 04$^{h}$43$^{m}$47$^{s}$.4 & $-$50$^{\circ}$16$'$59$''$ & $0.88^{+0.27}_{-0.13}$ & 9.0 & -1.7\\
  Reticulum 2 & 03$^{h}$35$^{m}$42$^{s}$.1 & $-$54$^{\circ}$02$'$57$''$ & $3.64^{+0.21}_{-0.12}$ & 7.0 & -2.3\\
  Reticulum III & 03$^{h}$45$^{m}$26$^{s}$.4 & $-$60$^{\circ}$27$'$00$''$ & $2.4^{+0.9}_{-0.8}$ & 7.0 & -1.2\\
  Sagittarius II & 19$^{h}$52$^{m}$40$^{s}$.5 & $-$22$^{\circ}$04$'$05$''$ & $2.0^{+0.4}_{-0.3}$ & 10.0 & -4.2\\
  Sculptor & 01$^{h}$00$^{m}$09$^{s}$.4 & $-$33$^{\circ}$42$'$33$''$ & $11.3^{+1.6}_{-1.6}$ & 5.0 & -3.3\\
  Segue II$^{\dagger}$ & 02$^{h}$19$^{m}$16$^{s}$.0 & +20$^{\circ}$10$'$31$''$ & $3.4^{+0.2}_{-0.2}$ & 12.0 & -1.8\\
  Sextans I & 10$^{h}$13$^{m}$03$^{s}$.0 & $-$01$^{\circ}$36$'$53$''$ & $27.8^{+1.2}_{-1.2}$ & 9.0 & -8.2\\
  Tucana 2 & 22$^{h}$51$^{m}$55$^{s}$.1 & $-$58$^{\circ}$34$'$08$''$ & $9.83^{+1.66}_{-1.11}$ & 8.0 & -3.1\\
  Tucana III*$^{\ddagger}$ & 23$^{h}$56$^{m}$36$^{s}$.0 & $-$59$^{\circ}$36$'$00$''$ & $6.0^{+0.8}_{-0.6}$ & 9.0 & -3.0\\
  Tucana IV*$^{\ddagger}$ & 00$^{h}$02$^{m}$55$^{s}$.2 & $-$60$^{\circ}$51$'$00$''$ & $9.1^{+1.7}_{-1.4}$ & 11.0 & -2.4\\
  Tucana V & 23$^{h}$37$^{m}$24$^{s}$.0 & $-$63$^{\circ}$16$'$12$''$ & $1.0^{+0.3}_{-0.3}$ & 8.0 & -3.0\\
\hline
\end{tabular}
\label{tab:dSphs}
\end{table*}

\subsection{Dwarf Spheroidal Galaxy Sample}
\label{sec:sample}

\begin{table}
    \centering
    \footnotesize
    \vspace{0.5cm}
    \caption{We show the basic imaging properties for two of the four GLEAM declination strips for which our sample of dSphs occupies. These statistics are derived from the wide-band (170\,--\,231\,MHz) image with values given as $mean\pm std.\,dev.$ \citep{Hurley-Walker2017}.}
    
    \begin{tabular}{ccccc} \hline\hline 
        \vspace{0.1cm}
        Property &  $-72^{\circ} < \delta < +18.^{\circ}5$ & $\delta > +18.^{\circ}5$ \\ \hline 
        Number of sources & 281,931 & 16,170 \\
        RMS ($\rm{mJy}\,\rm{beam}^{-1}$) & $10\pm5$ & $28\pm18$ \\
        PSF major axis ($'$) & $2.33\pm0.27$ & $3.20\pm0.23$ \\
        PSF minor axis ($'$) & $2.18\pm0.07$ & $2.25\pm0.03$ \\ \hline
        \label{tab:GLEAM_properties}
    \end{tabular} 
\end{table}

It has only been in the last decades that sensitive wide-field imaging surveys --- in particular, the Sloan Digital Sky Survey (SDSS; \citealt{York2000}) --- have allowed a class of ultra-faint dwarf galaxies to be detected as satellite companions to the Milky way and its neighbours. Examples of recently discovered faint galaxies include Bo$\ddot{\rm{o}}$tes II ($M_\mathrm{V} = -3.1$; \citealt{Walsh2007}), Coma Berenices ($M_\mathrm{V} = -3.7$; \citealt{Belokurov2007}), Segue 1 ($M_\mathrm{V} \sim -2$; \citealt{Geha2009}), Willman 1 ($M_\mathrm{V} = -2.5$; \citealt{Martin2007}).

We have formed a sample of Local Group dSph galaxies, listed in \citet{McConnachie2012} (as of September, 2015) and references therein, bar the addition of Crater 2, a newly discovered dwarf spheroidal galaxy candidate \citep{Torrealba2016}. The dSph galaxies in our sample mostly occupy the southern sky (although, at most, extend up to $\delta = +24^{\circ}$) and are at least 10$^{\circ}$ removed from the Galactic plane. We remove any galaxies that visually show obvious signs of having been disrupted. The basic optical properties and image qualities of the remaining dSph galaxies in the sample are given in Table \ref{tab:dSphs}. In this table, we note situations in which the GLEAM images of particular dSph galaxies suffer from imaging artefacts such as large-scale background gradients. This is most likely due to the sidelobes of nearby bright sources or a consequence of low-elevation observations. We also indicate galaxies that fall within $\sim3^{\circ}$ of each other. This is important when considering that the synchrotron signals from such pairs of galaxies may overlap with one another, thereby making it difficult to obtain independent measurements of their local backgrounds. Finally, there are two fields for which undeconvolved sidelobes of a nearby bright radio source dominate the image.

\noindent
\begin{figure}
    \centering
    \includegraphics[width=1\columnwidth]{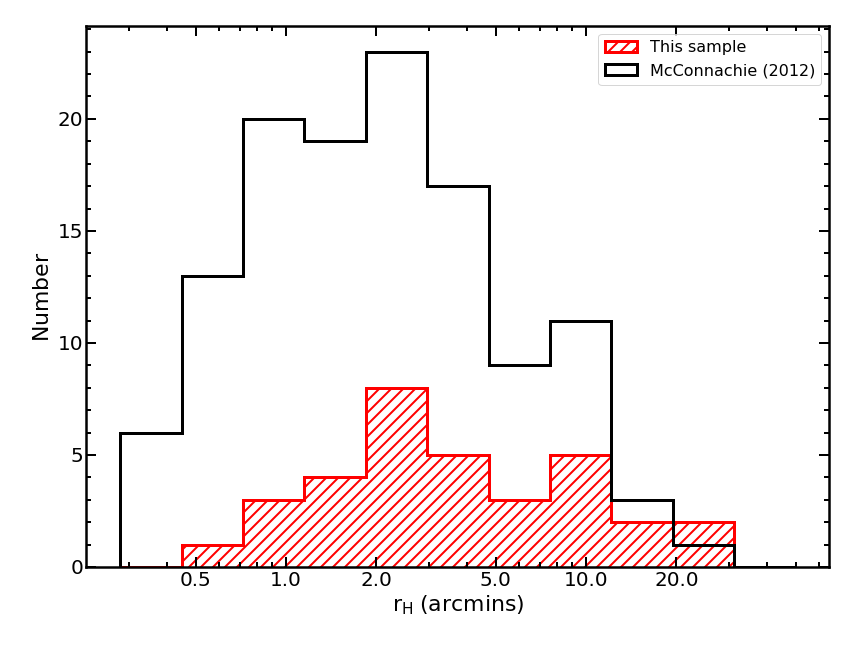}
    \caption{Histogram of the distribution of half-light radii for the sample of dwarf spheroidal galaxies used in our sample (red) as well as the full catalogue described in the \citep{McConnachie2012} (black).}
    \label{fig:dSphs_HL}
\end{figure}

Figure \ref{fig:dSphs_HL} shows the distribution of half-light radii for this sample of dSph galaxies with comparison to that of the full catalogue from \citet{McConnachie2012}. Roughly two thirds of our sample have half-light radii $\lesssim\,5'$. Given that the beam size of the MWA at $\nu =$ 170\,--\,231\,MHz is $\Omega_\mathrm{beam} = \frac{\pi}{4\ln(2)}\theta_\mathrm{maj}\theta_\mathrm{min} \sim 6.7\,\rm{square\,\,arcmin}$, for the majority of dSph galaxies, we are able to achieve $\gtrsim3$ independent measurements of the potential synchrotron emission within the half-light radius of the galaxy. However, for the 6 most compact dSph galaxies, we are restricted to a single independent measurement of the radio halo emission within the half-light radius.

\subsection{GLEAM Images}
\label{sec:GLEAM_images}
We have endeavoured to detect the signature of dark matter annihilations in the form of extended synchrotron emission from Local Group dwarf spheroidal galaxies with use of the GLEAM survey \citep{Wayth2015}. We extract $4^{\circ}\times4^{\circ}$ cutout images centred upon each dSph galaxy. We have utilised the wide-band (170\,--\,231\,MHz) stacked frequency images as these provide an equipoise of relatively high resolution and low image noise. These two factors should dictate our ability to detect dark matter annihilation signals. One requires a suitably high resolution such that discrete radio sources within the field can be accurately modelled and subtracted or masked, revealing the underlying brightness distribution intrinsic to the dSph galaxy.  Additionally, a low background noise is required in order for the inherently faint synchrotron signal to be detected. The imaging properties vary quite significantly across the GLEAM sky due, in part, to the projection of the MWA beam at different declinations. Table \ref{tab:GLEAM_properties} outlines the variation of the noise properties of the GLEAM survey for the declination ranges consequential to our sample.

\section{Image Processing}
\label{sec:method}
Here we address the exclusion of ten dSph fields from our sample due to various artefacts as well as provide a description of the image processing steps taken to produce radial brightness profiles for each of the dSph galaxies. We also describe the stacking of brightness profiles for both our dSph sample as well as for sets of randomly-selected control fields to determine the intrinsic scatter in our measurements.

\subsection{Problematic Fields}
\label{sec:problematic_fields}
Fields in the region of the sky bound by $10^{h} \lesssim \alpha \lesssim 22^{h}$ and $\delta \gtrsim +10^{\circ}$ have a typical RMS noise many times greater than fields outside of this region. Moreover, the variation of the RMS noise in these images appears to have a relatively steep gradient aligned with increasing declination. Given the $4^{\circ}\times4^{\circ}$ extent of our GLEAM cutout images, this results in a pronounced, degree-scale slope across the field. A feature such as this could be mistaken for the smoothly distributed, degree-scale synchrotron emission signal that we are trying to extract. Whilst we weight the measurements according to the inverse square of the average RMS within each annulus, we have opted to remove these particular dSph galaxies from our sample as we know that any apparent signal present in their radial brightness profiles is likely to be extrinsic to the galaxy. 

In our sample, there are two adjacent fields in which a bright source contaminates the image to such a degree that any synchrotron emission that may be present from the galaxy is likely washed out by the sidelobes of the bright source. The bright radio source (PKS 2356-61; $\alpha = $23$^{h}$59$^{m}$04.3$^{s}$, $\delta = $ -60$^{\circ}$54$'$59$''$) affects both Tucana III and Tucana IV, hence we remove both from the analysis.

Each of the radial brightness profiles is normalised to the background level of the region of sky in which it is located. In cases where two dSph galaxies are within $\sim3^{\circ}$ of each other, this poses the problem of potentially measuring the background level of the field with the additional contribution of the intrinsic synchrotron emission from the neighbouring galaxy. It is difficult to impose a reasonable overlap tolerance as the extent of the dark matter halo is not truly known. As it happens, the three such examples of overlapping pairs (see Table \ref{tab:dSphs}) exhibit other shortcomings in their fields, be that steep gradients in the background emission or poorly subtracted bright sources.

\subsection{Background Source Subtraction}
\label{sec:source_subtraction}
The GLEAM extragalactic catalogue contains $\sim3\times10^{5}$ discrete radio sources which are predominantly unresolved at 170\,--\,231\,MHz. The apparent shapes of these point-like sources in the survey will be determined by the point spread function (PSF). Most sources can be modelled accurately as either a single Gaussian component or as a collection of these, described by the source's peak flux density, its major and minor axes and position angle.

\noindent
\begin{figure}
    \centering
    \vspace{0.25cm}
    \includegraphics[width=1\columnwidth]{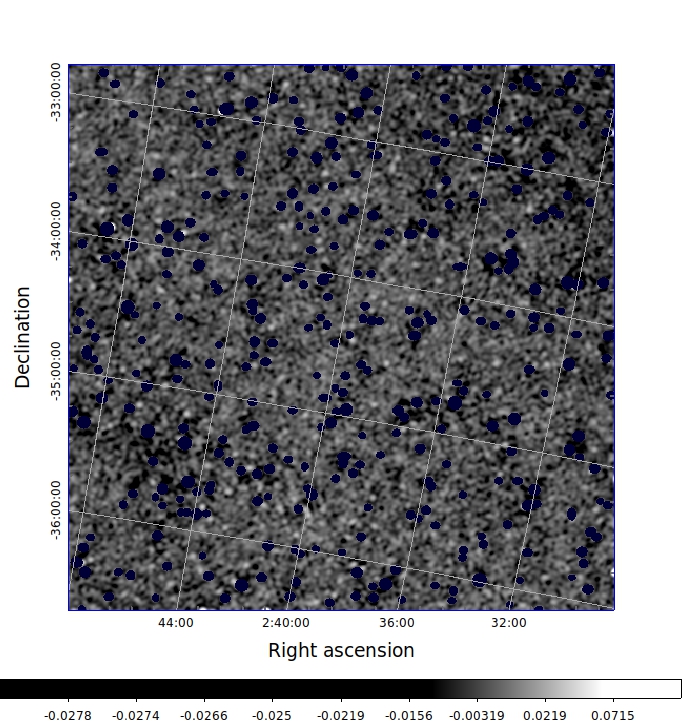}
    \caption{An example of the background radio source masking with a $4^{\circ}\times4^{\circ}$ GLEAM image of the Fornax dSph field in the 170\,--\,231\,MHz frequency band. Elliptical masks have been applied by masking only the pixels corresponding to the sources listed in the GLEAM extragalactic catalogue \citep{Hurley-Walker2017}. The linear brightness scale ranges from $-28$\,--\,$170$\,\mJybeam. The image has a synthesised beam of width 2.28$'$ and a sensitivity of $\sigma_\mathrm{RMS} = 5.5$\,\mJybeam. The dark blue ellipses show the masked regions of background radio sources.}
    \label{fig:Fornax_residual}
\end{figure}

In order to detect the \emph{extended} synchrotron emission signal from the radio halo of a dSph galaxy, we first mask the background sources from the GLEAM extragalactic catalogue \citep{Hurley-Walker2017} that fall within each dSph field. The number of sources in a given $4^{\circ}\times4^{\circ}$ field is typically of order $10^{2\,-\,3}$, depending on the region of the GLEAM survey. Since the image response to a radio source is the convolution of its true shape with the beam, the vast majority of unresolved background sources can be well modelled as a 2-dimensional Gaussian ellipse. We utilise \textsc{Aegean} Residuals \citep[\textsc{AeRes} ][\footnote{\href{https://github.com/PaulHancock/Aegean}{https://github.com/PaulHancock/Aegean}}]{Hancock2018} to perform the source extraction process. We define elliptical masks centred upon each source with side lengths corresponding to $1.5\times$ its measured semi-major and semi-minor axes. As an example, Figure \ref{fig:Fornax_residual} shows this masking process for the Fornax dSph galaxy. In our source-subtracted images, we typically achieve a sensitivity of $\sigma_\mathrm{sub} \lesssim 10$\,\mJybeam. It is from these masked images that we attempt to extract the extended synchrotron emission signal. There are, of course, remaining background radio sources within the images that have fallen below the detection threshold of the GLEAM survey (see Table \ref{tab:GLEAM_properties}) that are not masked in this process. However, we assume these to have a fairly uniform distribution across the image, causing only a relative increase in the background level across the image \citep{Spekkens2014}.

\begin{figure*}
    \centering
    \includegraphics[width=0.65\textwidth]{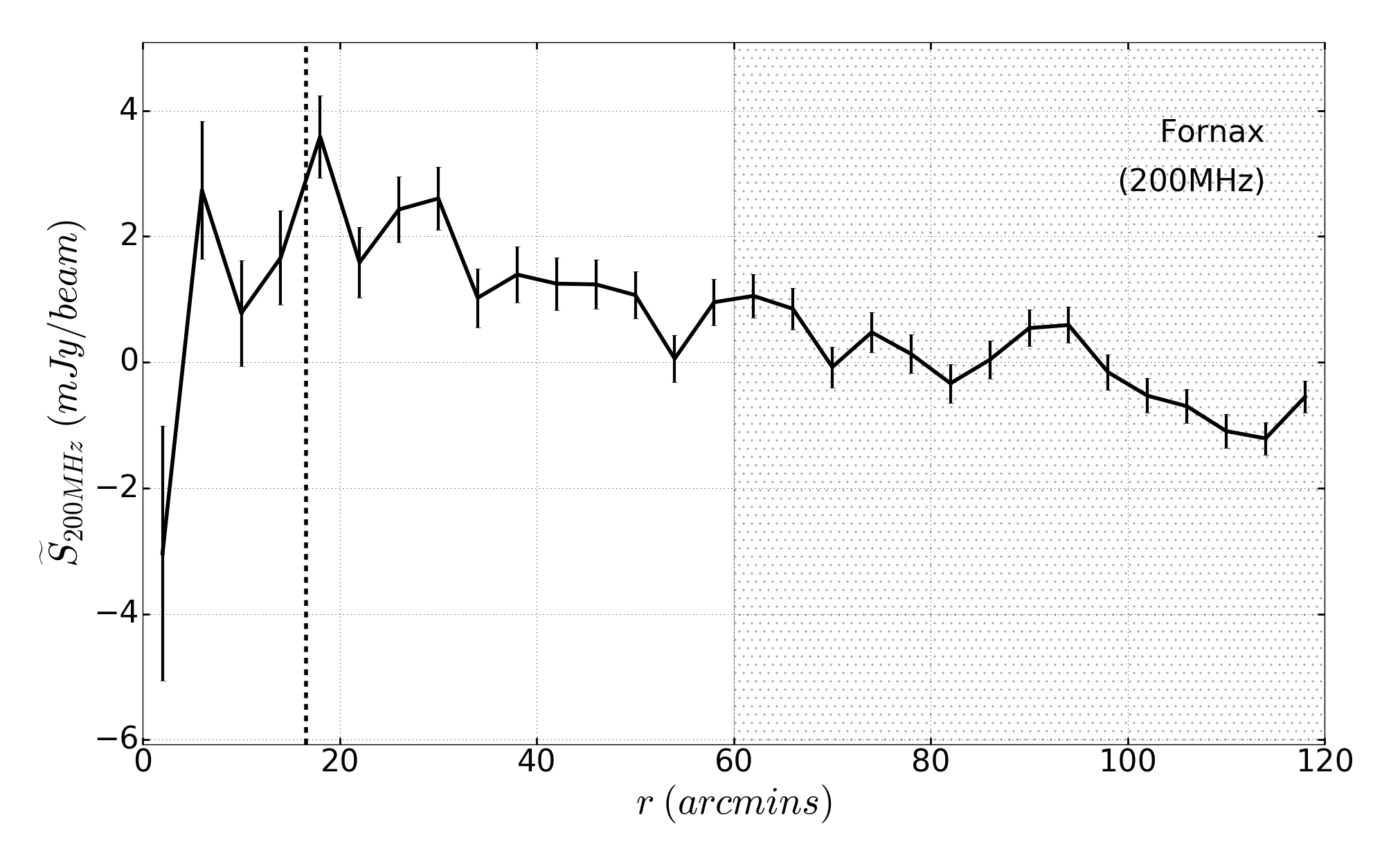}
    \caption{Radial brightness profile for the Fornax dSph galaxy using a $4^{\circ}\times4^{\circ}$ GLEAM wide-band image (170\,--\,231\,MHz). The median flux measurements are given in units of \mJybeam and have been normalised to the background level of the field, measured in the hatched region. The semi-major axis of the beam is $\sim2.3'$. The dashed vertical line indicates the half-light radius of Fornax.}
    \label{fig:FLvR}
\end{figure*}
 
Moreover, the masking of background sources may also be imperfect, predominantly due to inaccuracies in the measured parameters of the Gaussian ellipsoids. Sources --- and in particular, clusters of sources --- that have not been measured correctly may result in an excess of flux beyond the masking region. For this reason, we apply an iterative sigma clipping routine to the residual images to further mask all pixels with flux density values outside a $\pm2.5\sigma$ threshold of the local RMS noise. Compared to a method in which models of each source are subtracted from the image, this method retains fewer pixels for stacking, but the contribution from bright background radio sources in the GLEAM extragalactic catalogue is negligible. Directly subtracting models of the sources from the images often leads to spurious residuals due to the deviations that real sources have from the simplistic Gaussian ellipse model.

\subsection{Radial Surface Brightness Profiles}
\label{sec:FLvR}

Once we derive residual maps for each of our dwarf spheroidal galaxies, we are in a position to determine their radial surface brightness profiles. Each image is partitioned into radial bins bound by concentric annuli with $4'$ widths. This particular width was chosen to balance the need to probe the innermost regions of the dwarf galaxy as finely as possible against our ability to make individual independent measurements with the MWA beam (see Table \ref{tab:GLEAM_properties}). Bins in the outer regions of the galaxy contain more pixels than those towards the centre of the field (by as much as two orders of magnitude). Each average flux density measurement has an associated uncertainty, given by $\sigma_\mathrm{RMS}/\sqrt{N}$, where $N$ is the number of beams within an annulus; it follows that the largest uncertainties will be in the innermost radial bins. The RMS noise, $\sigma_\mathrm{RMS}$, within each image was determined using \textsc{Bane}, an algorithm as part of \textsc{Aegean} \citep{Hancock2012}.

We compute the background level for each dSph galaxy field as the mean flux density in the region between $60'$\,--\,$120'$. The background level is then subtracted from each bin such that the brightness profile is normalised about zero. The median flux density is then remeasured within each annulus and plotted as a function of radius to generate the radial brightness profile for the galaxy. As an example, Figure \ref{fig:FLvR} presents the radial brightness profiles for the Fornax dSph galaxy --- one of the more extended dSph galaxies in the sample --- using a wide-band (170\,--\,231\,MHz) GLEAM image. The radial surface brightness profiles for all 33 dSph galaxies can be seen in Figure \ref{fig:galaxy_profiles}.

\subsection{Stacking Profiles}
\label{sec:stacking}
Since the surface brightnesses of individual dwarf spheroidal galaxies are expected to be quite faint at radio frequencies (cf. \citealt{Spekkens2013,Regis2015}), we have opted to stack the radial brightness profiles in an attempt to obtain a statistical detection of a possible synchrotron imprint left behind by the dark matter particle annihilation products.

For each radial bin, we take the average of the median flux measurements over all dSph galaxies in the sample, weighted by $1/\sigma_\mathrm{\rm{RMS}}^{2}$, where $\sigma_\mathrm{RMS}$ is the root mean square noise within an annulus for a particular galaxy. Figure \ref{fig:dSphs_HL} illustrates that our sample contains dSph galaxies with a wide range of effective radii. For this reason, we also attempt to stack the radial brightness profiles in physical units by scaling each dSph galaxy radially to its respective half-light radius. As the half-light radii vary over a wide range of angular sizes (Table \ref{tab:dSphs}), the profiles of largest galaxies are confined to within a few physical radial bins whereas a single independent measurement from the smallest galaxies may span several bins. The measurements of the half-light radii for these galaxies have quite large uncertainties which makes the analysis of the stacking in physical units more difficult. For completeness, we also stacked the dSph galaxies within the image plane first as opposed to stacking the radial brightness profiles directly. We find little difference between these two methods.

\begin{figure}
    \centering
    \includegraphics[width=0.47\textwidth]{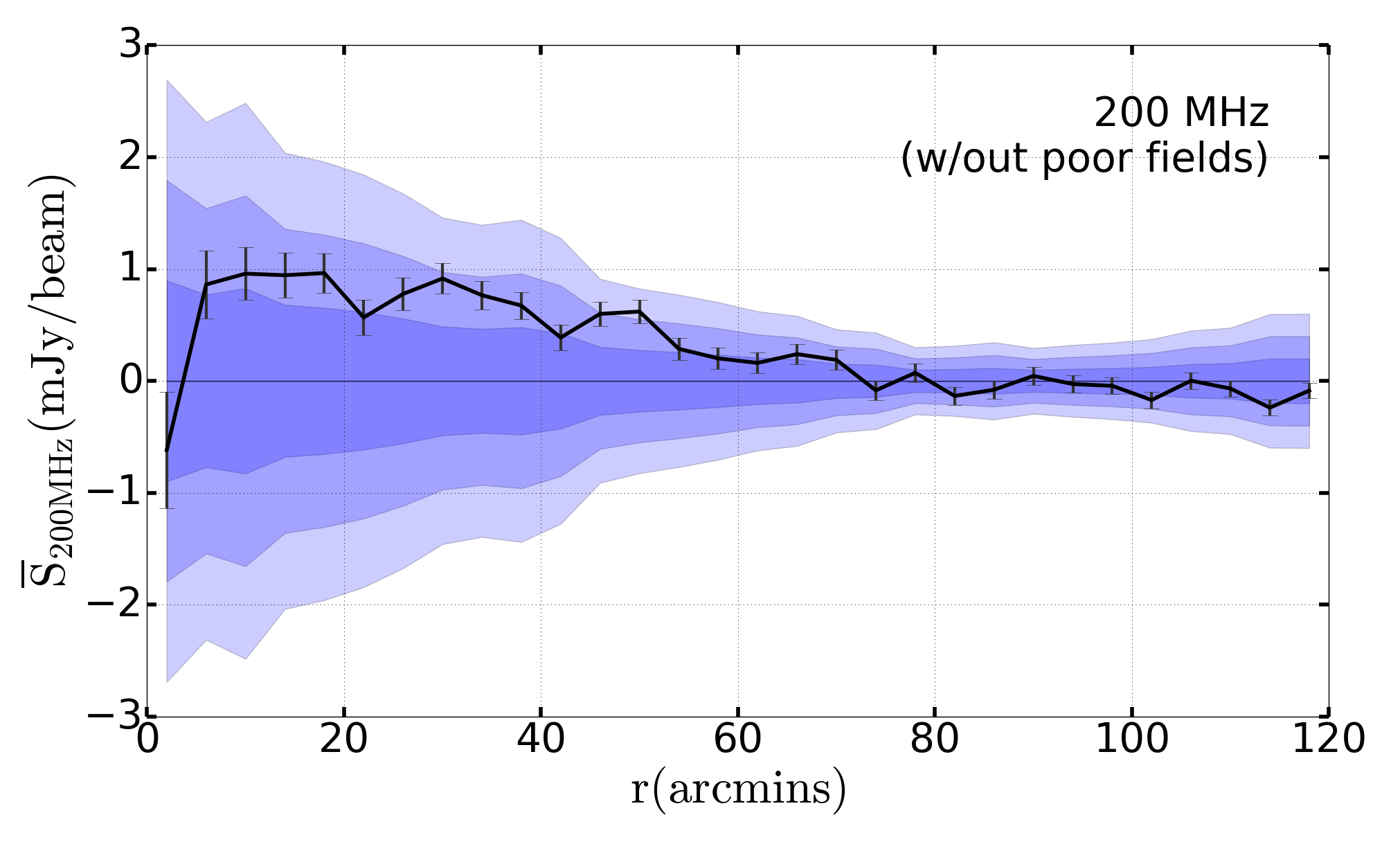}
    \includegraphics[width=0.47\textwidth]{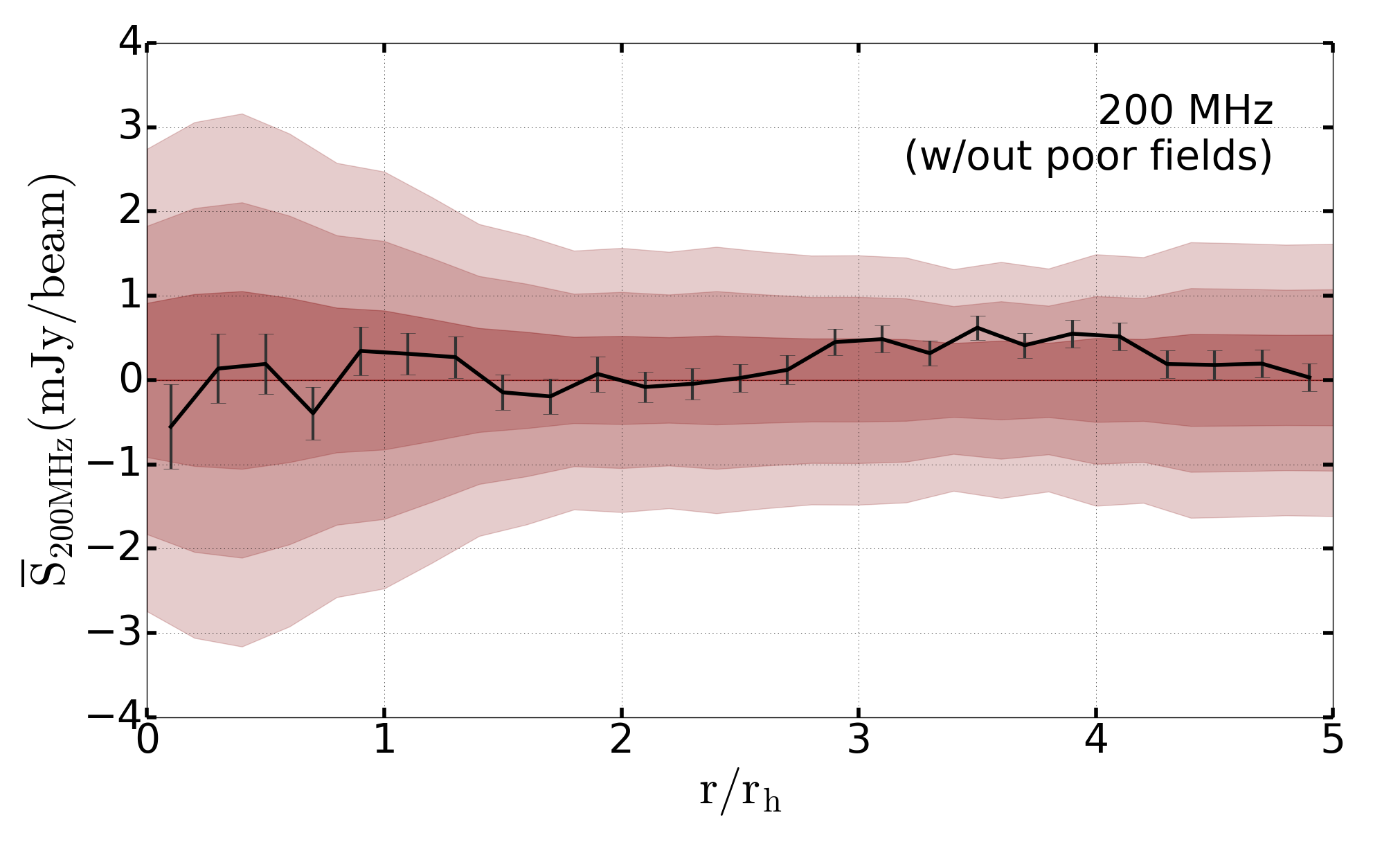}
    \caption{Radial brightness profile stacking of the 23 dSph galaxies viable for stacking in (top) angular radial bins and (bottom) scaled to the half-light radius of each galaxy. The flux density measurements are the average of the median flux density measurements within each radial bin over all galaxies, weighted by the inverse square of the mean RMS of each bin. The shaded regions show the $\sigma = 1,2,3$ RMS noise levels for the stacking of 1000 radial brightness profiles from randomly selected control fields in representative regions of the sky.}
    \label{fig:voids_stacked}
\end{figure}

\subsection{Control Fields}
\label{sec:controls}
Our sample is relatively small, resulting in large fluctuations in the stacked profiles. In order to quantify the significance of the stacked galaxy radial profiles, we perform exactly the same stacking analysis on a large set of randomly chosen control fields. Because of the large-scale variation in background emission present across the GLEAM sky, it is necessary for our control fields to be located in similar regions of the sky to the target dSph galaxies. Each target dSph galaxy was assigned six adjacent control fields unless --- as is often the case --- the field overlapped with a neighbouring dSph galaxy. In this case, the control field was assigned to the next closest sky position which met this requirement. This process ensures that the control fields remain representative of the original sky statistics.

For each of the control fields, we follow the same procedure as described in Sections \ref{sec:source_subtraction} and \ref{sec:FLvR} to generate a masked image and radial surface brightness profile. We then produce $N = 10^{3}$ permutations of 23 control fields (one for each galaxy) with the condition that no two control fields within a single permutation should redundantly overlap. The stacking algorithm is performed on all permutations as above. Within each radial bin, we derive the RMS variation from all $N$ measurements of the median flux which can be thought of as the intrinsic spread of the radial brightness profiles for regions of the sky in which no signal is expected. Comparing this against the stacked profile for the fields containing the dSph galaxies will give an indication of the significance of any features found therein.

\section{RESULTS}
\label{sec:results}
Here we present the results of the stacking analysis for our sample of dwarf spheroidal galaxies in both apparent sky coordinates as well as scaled by the half-light radius of each dSph galaxy. Figure \ref{fig:voids_stacked} presents the result of the stacking the 23 galaxies in the sample that were deemed viable for stacking (see Table \ref{tab:dSphs}). The shaded regions show the intrinsic scatter in the light profiles of control fields at the levels of 1, 2 and 3 $\sigma$.

\begin{figure*}
    \centering
    \includegraphics[width=0.47\textwidth]{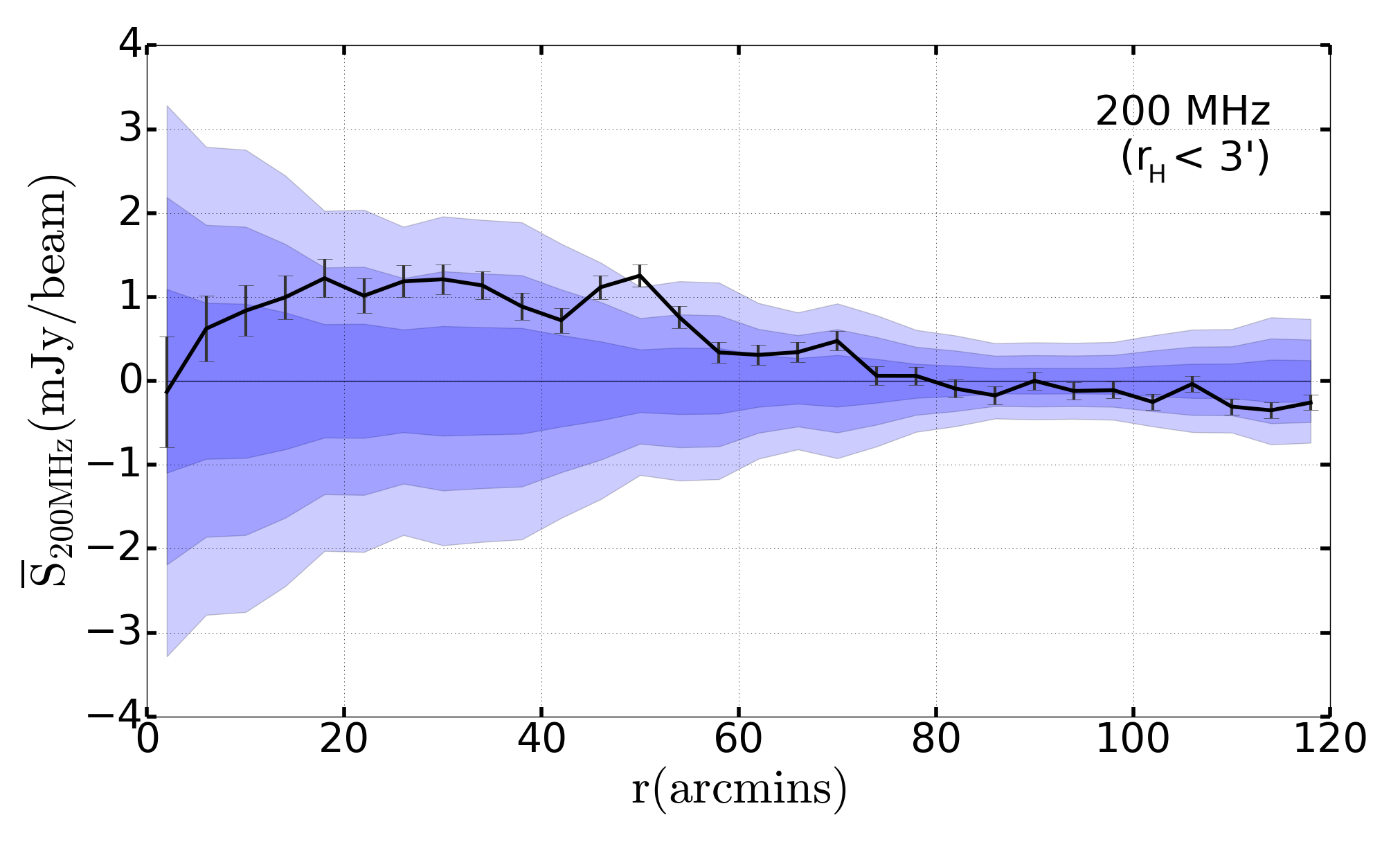}
    \includegraphics[width=0.47\textwidth]{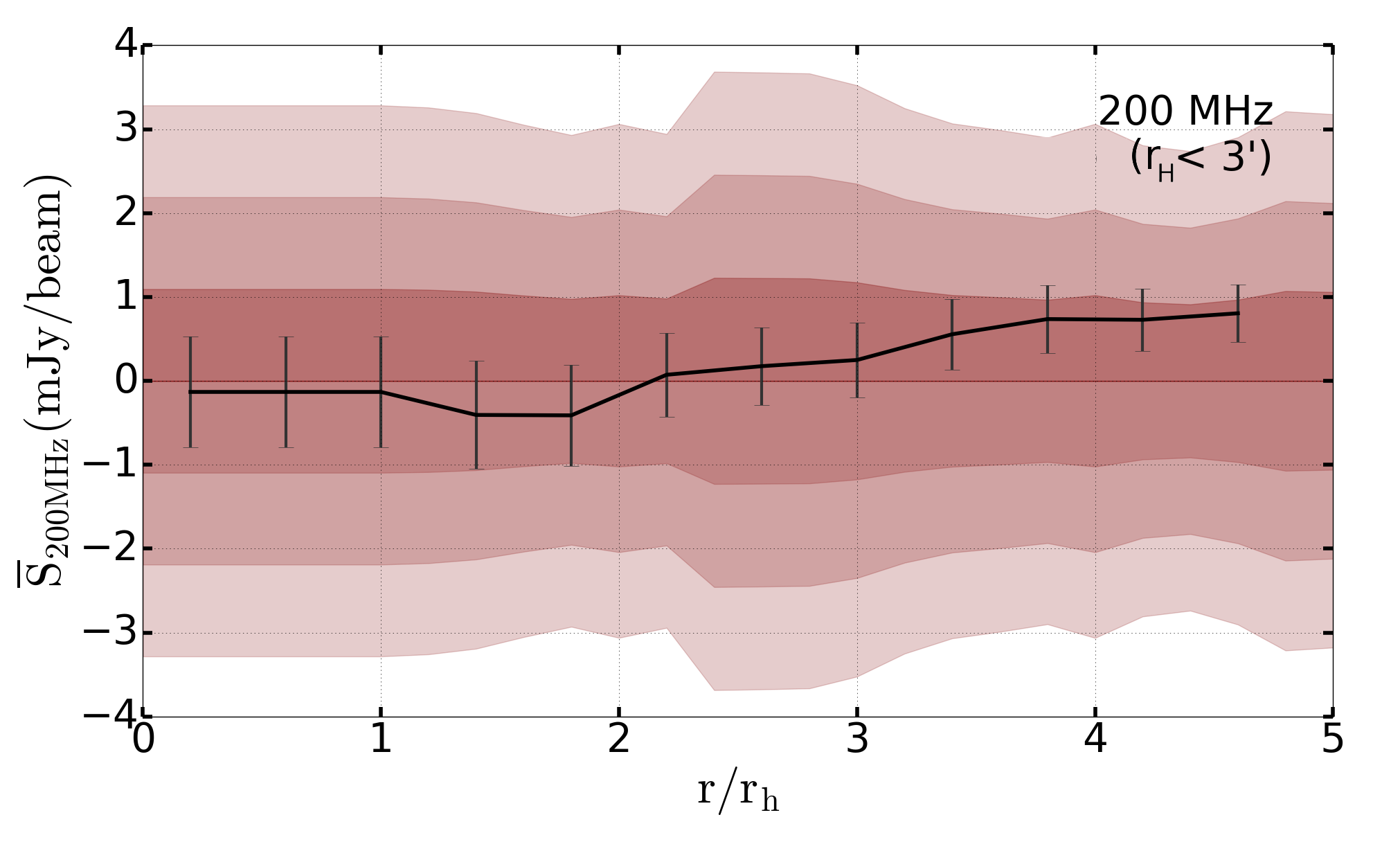}
    \caption{Radial brightness profile stacking at 200\,MHz for dSph galaxies with half-light radii $r_\mathrm{h} < 3'$. Discounting the removed poor fields (see Table \ref{tab:dSphs}), this subset contains 14 dSph galaxies. Stacking is performed in angular radial bins (left) as well as scaling according to the respective half-light radii (right). Background source masking has been performed with additional sigma clipping.}
    \label{fig:below_3arcmin}
\end{figure*}

\begin{figure*}
    \centering
    \includegraphics[width=0.48\textwidth]{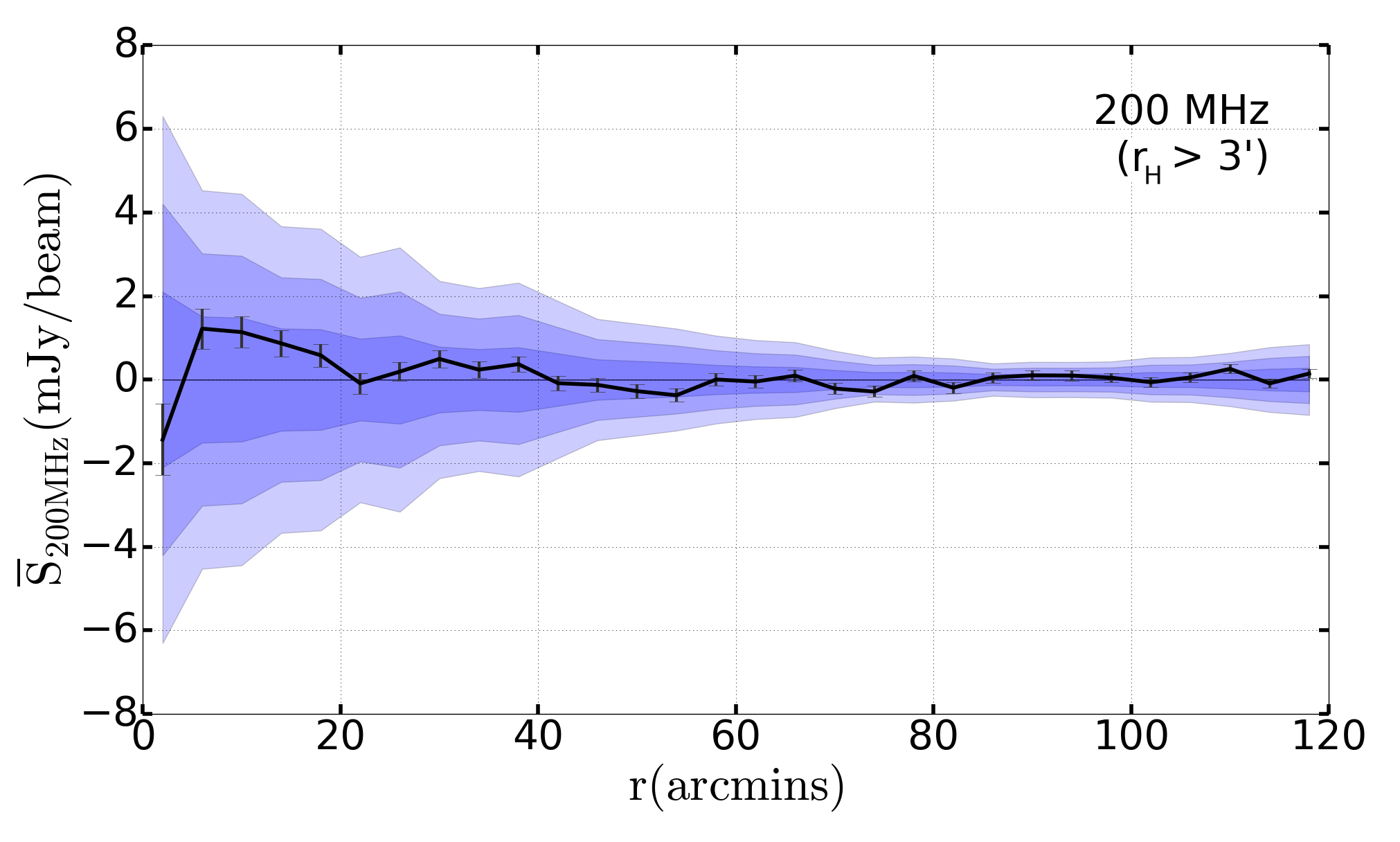}
    \includegraphics[width=0.48\textwidth]{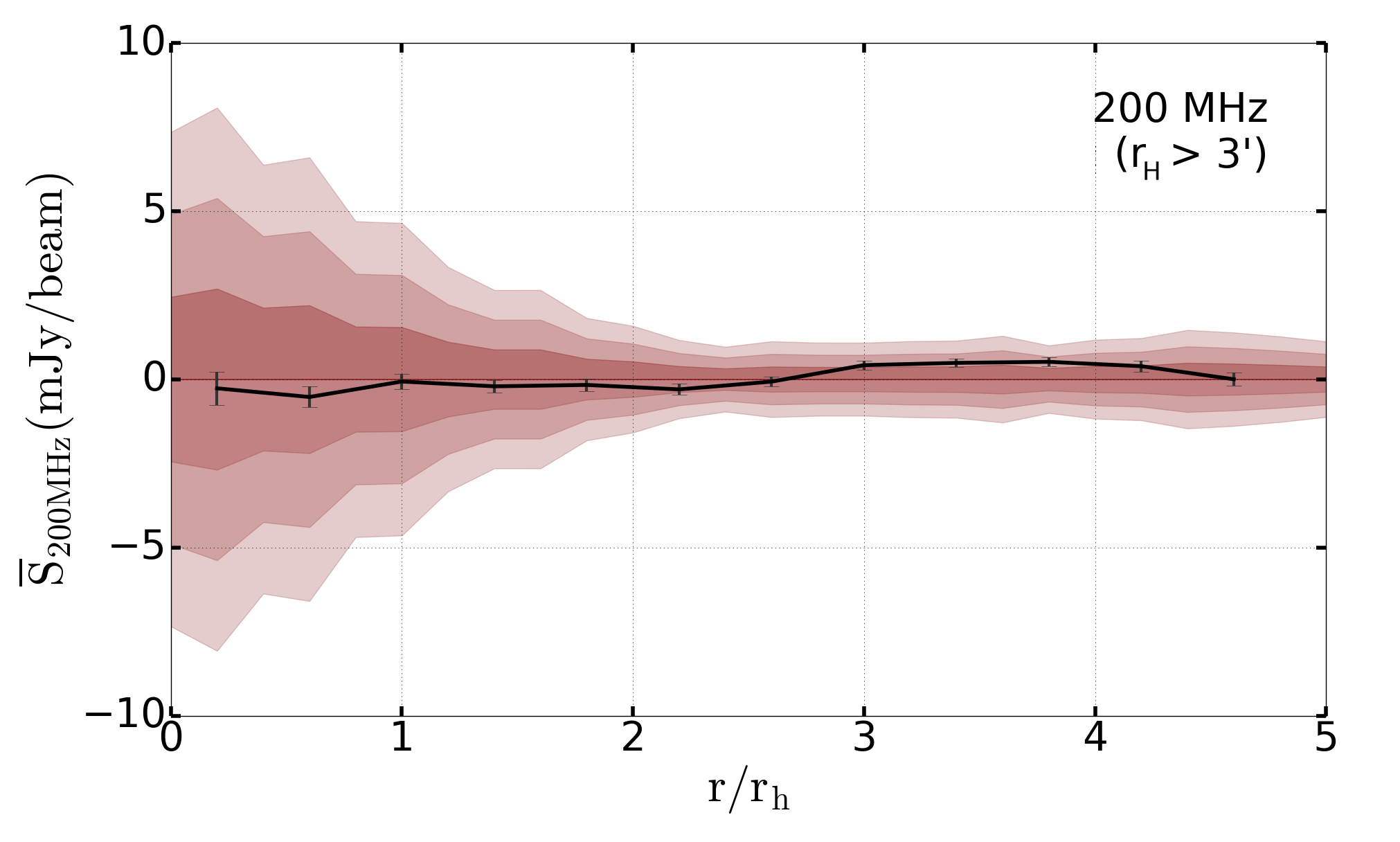}
    \caption{Radial brightness profile stacking at 200\,MHz for dSph galaxies with half-light radii $r_\mathrm{h} > 3'$. Discounting the removed poor fields (see Table \ref{tab:dSphs}), this subset contains 9 dSph galaxies. Panel descriptions are as above.}
    \label{fig:above_3arcmin}
\end{figure*}

\citet{Colafrancesco2007} model the distribution of charged electron and positron pairs by noting that there is a random component of interstellar magnetic fields associated with the dSph galaxy systems and that it is a fair approximation to model the propagation of charged particles as a diffusive process. In the case of the Draco dSph galaxy, the region of diffusion extends approximately twice as far as its stellar component. The sample of dwarf spheroidal galaxies used in this analysis have a large range of effective radii (see Figure \ref{fig:dSphs_HL}). To a first approximation, we estimate the extent of the annihilation signal for the dSph galaxies in this sample by scaling according to their half-light radius. This means that for more extended dSph galaxies, we would expect a more diffuse propagation of electrons and positrons in their halos. This will cause the synchrotron signal to be fainter as it will be dispersed over a larger area.

We have plotted realisations of the stacking analysis for subsets of the sample with half-light radii below and above a $r_{H} = 3'$ cut in Figures \ref{fig:below_3arcmin} and \ref{fig:above_3arcmin}, respectively. The stacking of dSph galaxies with half-light radii less than 3$'$ in physical units becomes noisy as we are limited to very few independent measurements of the flux density in the given range. This means that the stacking analysis generally contains only the inner radial bins which have the greatest uncertainties in their measurements.

\section{DISCUSSION}
\label{sec:discussion}

\subsection{Stacked Radial Brightness Profiles}
The results of stacking the sample of dwarf spheroidal galaxies (Figure \ref{fig:voids_stacked}) shows a positive signal at a 1.5$\sigma$ level across the innermost $60'$, which we conclude to be consistent with a null detection. The 95\% confidence interval ($\sim 2\sigma$) of the intrinsic random scatter peaks at a flux of 1\,--\,2\,\mJybeam within the innermost radial bins, depending on whether the images are scaled to their half-light radius or not. Given that we do not see a detection at this level, this upper limit could provide constraints on the possible cross-sections for potential dark matter candidates. Such values are model-dependent and crucially dependent upon the magnetic field strengths ($\mathrm{B}$), the degree of spatial diffusion ($D_{0}$) in each dSph galaxy and the particular dark matter density profile ($\rho_{\chi}(r)$) assumed \citep{Colafrancesco2006,Colafrancesco2007}. In particular, the magnetic field strengths of dSphs are not well constrained as polarisation measurements are difficult to obtain due to a low gas and dust content \citep{Spekkens2013,Natarajan2013}. It is typical that star-forming dwarf galaxies of the Local Group have magnetic field strengths of order $\sim\mu\mathrm{G}$ \citep{Chyzy2011}. However, the magnetic fields of galaxies affects the signal strength both from the relation giving the total radiative power of synchrotron emission as well as due to losses from radiative processes. Thus, providing quantitative constraints on dark matter properties is beyond the scope of this work, as such we do not supply these upper limits here.

However, we are able to make a comparison to the work of \citet{Spekkens2013} who estimate the radial surface brightness profile of the Draco dSph for observations with the Green Bank Telescope (GBT) at $\nu_\mathrm{\textsc{gbt}} = 1.4\,\rm{GHz}$ based upon a series of models by \citet{Colafrancesco2007}. These models assume a WIMP mass of $M_\mathrm{\chi} = 100\,\rm{GeV}$ annihilating into \textit{$b\overline{b}$} within a turbulent magnetic field strength in Draco of $B = 1\,\mu G$ and with annihilation cross-section $\langle\sigma v\rangle_\mathrm{\chi} = 3.4\times10^{-23}\,\rm{cm}^{3}\,\rm{s}^{-1}$. This leads to a prediction in the peak intensity of order 1 \Jybeam at the innermost region of the galaxy. Note that this value of $\langle\sigma v\rangle_\mathrm{\chi}$ is now strongly ruled out by $\gamma$-ray searches. However, the predicted intensity may be scaled linearly to a more realistic value of $\langle\sigma v\rangle_\mathrm{\chi} \approx 3\times10^{-26}\,\rm{cm}^{3}\,\rm{s}^{-1}$ as determined by the thermal relic on the observed dark matter density of the Universe \citep{Porter2011}. Accounting for the fact that the GBT beam is $\theta_\mathrm{\textsc{gbt}} = 9.12'$ (compared to $\theta_\mathrm{\textsc{mwa}} \sim2.5'$ for the MWA) and scaling to a frequency of $\nu_\mathrm{\textsc{mwa}} = 200$\,MHz assuming a spectral index of $\alpha = -0.7$, we would expect to observe a peak intensity of order $\sim$1\,\mJybeam for the Draco dSph. This assumed spectral index is derived by adopting Milky Way-like diffusion models describing charged particle propagation in the form of $D(E) \propto D_{0}E^{\gamma}$, with $D_{0} = 3 \times 10^{26}\;\mathrm{cm^{2} s^{-1}}$ and $\gamma = -0.6$ (``set \#2'' in Fig. 12 of \citealt{Colafrancesco2007}). This prediction is quantitatively similar to the constraints obtained by \citep{Spekkens2013} using the GBT.

It has become common amongst previous literature to describe the dependence of the dark matter density distribution on a galaxy's total annihilation rates using the so-called J-Factor. This factor is the line-of-sight integral of the dark matter density over a given solid angle and can be estimated through the measured stellar kinematics of dSphs. Recently, \citet{Pace2019,Evans2016} have computed the J-Factors of dSphs with measured stellar kinematics and further calculated estimates for systems where these measurements are not available \citep{Pace2019}. For the sample of dSph galaxies used in this stacking analysis, the average J-Factor for those with measurements available is 18 GeV$^{2}$ cm$^{-5}$ measured within an angular cone of $\theta_\mathrm{max}= 0.5^{\circ}$. Using more sensitive radio data, the resulting constraints on dark matter properties using this stacking technique will be highly complementary to those obtained from $\gamma$-ray experiments \citep{Ackermann2011,Albert2017,Hoof2018}.

Further to this, we have found that stacking in terms of radial bins scaled to the half-light radii of each galaxy does not produce a stronger signal. This is likely a result of the wide range of half-light radii (and associated uncertainties) in our sample which causes the smallest dSph galaxies to be severely under-sampled in the inner regions of the image. This may also explain the jagged appearance of the mean RMS fluctuations of half-light bins in the sample of random control fields which are also scaled to the half-light radius of the corresponding galaxies.

We note that the scaling process relies upon the assumption that the extent of the dark matter halo scales with the effective optical radius of the dSph galaxy. In order for this to be the case, there is the requirement that each dSph galaxy has a fixed annihilation flux which depend upon their specific stellar velocity distributions. Models by \citet{Colafrancesco2007} show that in the case of the Draco dSph (not visible with the MWA), its surface brightness is expected to drop by an order of magnitude at a radius of $\sim3\,r_\mathrm{h}$. Furthermore, \citet{McConnachie2012} and references therein, show that there are large uncertainties associated with measurements of the half-light radius for the dSph galaxies in our sample. Typically, they have a fractional uncertainty of $\sim15$\% in the half-light radius, but can be as large as 50\% in the case of the more compact systems.

\subsection{Implications for Future Radio Continuum Surveys}
In our analysis, several limitations have lessened our ability to detect the expected synchrotron emission signal from the halos of dwarf spheroidal galaxies: low sensitivity observations, limited capacity to probe the inner regions of galaxies, and a poor measure for the Galactic foreground emission in our images. The GLEAM data used in this analysis provide relatively shallow snapshot imagery with exposure times of $\tau_\mathrm{eff}\sim10$\;minutes. If we trust, impetuously, the $\sim1.5\,\sigma$ detection from the stacking of galaxies in our sample (Figure \ref{fig:voids_stacked}), we can estimate by how much longer we would need to observe in order to obtain a $5\sigma$ detection. As the signal-to-noise ratio of our measurements goes as the square of the exposure time, we estimate that an exposure time of $\tau = (5/1.5)^{2}\times 10\,\rm{[min]} \sim 2\,\rm{hrs}$ for each dSph galaxy would be required.

Such depths are now possible with the MWA phase 2 \citep{Wayth2018} which includes the addition of new tiles forming baselines up to $\sim 5.5\,$km, effectively doubling the resolution and decreasing both the natural and sidelobe confusion by more than an order of magnitude. Currently on-going is GLEAM-eXtended (GLEAM-X; Hurley-Walker et al. in prep.), a deeper higher resolution version of GLEAM, as well as several very deep pointings (Seymour et al. in prep.) of well-studied extra-galactic fields coinciding with the Galaxy And Mass Assembly survey (GAMA; \citealt{Driver2011, Liske2015}). Notwithstanding the improved sensitivity to extended emission, it was recognised by \citet{Spekkens2013} that emission from the Galactic foreground itself becomes a major limitation. In future work, it should also be possible --- based on the \citet{Colafrancesco2007} models --- to estimate how large of a background sky box is required in order to adequately model and subtract foreground emission whilst avoiding the subtraction of emission from the sources themselves. Furthermore, our use of the RMS noise as a measure of the distribution of noise pixels across the images may bias the analysis as the noise is not completely Gaussian. In the case where classical confusion is the dominant noise source, the sky will have a real positive background distribution which is likely to contaminate the faint dark matter signal.

Some of these limitations could be addressed using observations from facilities such as the Australia Telescope Compact Array (ATCA). We estimate that with an observing frequency of 2.1\,GHz in the compact H214 configuration of ATCA, we would be able to achieve a brightness sensitivity as low as $\sim$0.3\,mK; the resolution of the synthesised beam for this particular observational setup would be $\sim1.5'$. For long exposure times, the increased brightness sensitivity would cause the observations to be dominated by the contributions of faint sources in the field. These discrete sources could be observed, modelled and subtracted with higher resolution observations either with ATCA or the Australian Square Kilometre Array Pathfinder (ASKAP). The Evolutionary Map of the Universe (EMU; \citealt{Norris2011}) project will observe the entire southern sky at $\sim 1\,$GHz with a resolution of $10''$, but with enough short baselines to be sensitive to emission on scales of order several degrees. In this way, confused radio sources can be subtracted whilst remaining sensitive to potential synchrotron emission from dSph galaxies. These future observations --- in addition to those forthcoming from the SKA ---  will prove to be very powerful for probing particle dark matter annihilation models by providing constraints on the annihilation cross-section that are several orders of magnitude more stringent than previous $\gamma$-ray observations with Fermi-LAT \citep{Colafrancesco2015}.

\section{CONCLUSIONS}
\label{sec:conclusions}
We present a search for smoothly distributed, degree-scale radio emission in a sample of 23 dwarf spheroidal galaxies with observations at $\sim$200\,MHz from the Galactic and Extragalactic All-sky Murchison Widefield Array (GLEAM) survey in an attempt to detect a dark matter annihilation signature. We extract $4^{\circ}\times4^{\circ}$ wideband frequency (170\,--\,231\,MHz) images centred upon each of the dSph galaxies in our sample with declination $\delta \lesssim 30^{\circ}$ and sensitivities ranging between $\sigma\sim$\,5\,--\,15\,\mJybeam. With an adaptation of \textsc{Aegean}, we mask discrete sources from the GLEAM extragalactic catalogue within these images to produce residual maps, which are further sigma clipped to a level of 2.5\,$\times$\,sigma such that artefacts caused by the masking process do not contaminate the field. For each dSph galaxy field, we compute a radial surface brightness profile from the azimuthal averaging of the flux density within concentric annulus bins, out to a radius of $2^{\circ}$. Each radial brightness profile is normalised by subtracting away the mean background level in the region defined as $60' < r < 120'$ from the field centre.

We stack the individual brightness profiles in an attempt to obtain a statistical detection of the synchrotron emission. The stacking analysis is performed both in terms of angular units as well as scaled radially according to the half-light radius of each dSph galaxy. The stacked signal is at a 1.5$\sigma$ significance within $r < 60'$ when stacked in the sky plane. Hence, we do not detect a statistically significant signal due to synchrotron emission, consistent with a null detection. Stacking each dSph galaxy scaled in units of their half-light radius produced a less significant result; the predicted smoothly-declining surface brightness profile is therefore not observed. This novel technique illustrates the capacity for using low-frequency radio images to test theories of particle dark matter in future low radio frequency continuum surveys.

\section*{Acknowledgements}
\small
The authors would like to thank the reviewer for their insightful comments which significantly improved the quality of this work.\\

This scientific work makes use of the Murchison Radio-astronomy Observatory, operated by CSIRO. We acknowledge the Wajarri Yamatji people as the traditional owners of the Observatory site. Support for the operation of the MWA is provided by the Australian Government (NCRIS), under a contract to Curtin University administered by Astronomy Australia Limited. We acknowledge the Pawsey Supercomputing Centre which is supported by the Western Australian and Australian Governments.

This research made use of ds9, a tool for data visualization supported by the Chandra X-ray Science Center (CXC) and the High Energy Astrophysics Science Archive Center (HEASARC) with support from the JWST Mission office at the Space Telescope Science Institute for 3D visualization. 

\normalsize
\addcontentsline{toc}{section}{Acknowledgements}



\bibliographystyle{mnras}
\bibliography{bibfile} 


\appendix
\section{Radial Brightness Profiles}
\label{sec:appendix_radial_brightness_profiles}

\begin{figure*}
    \centering
    \includegraphics[width=0.9\textwidth]{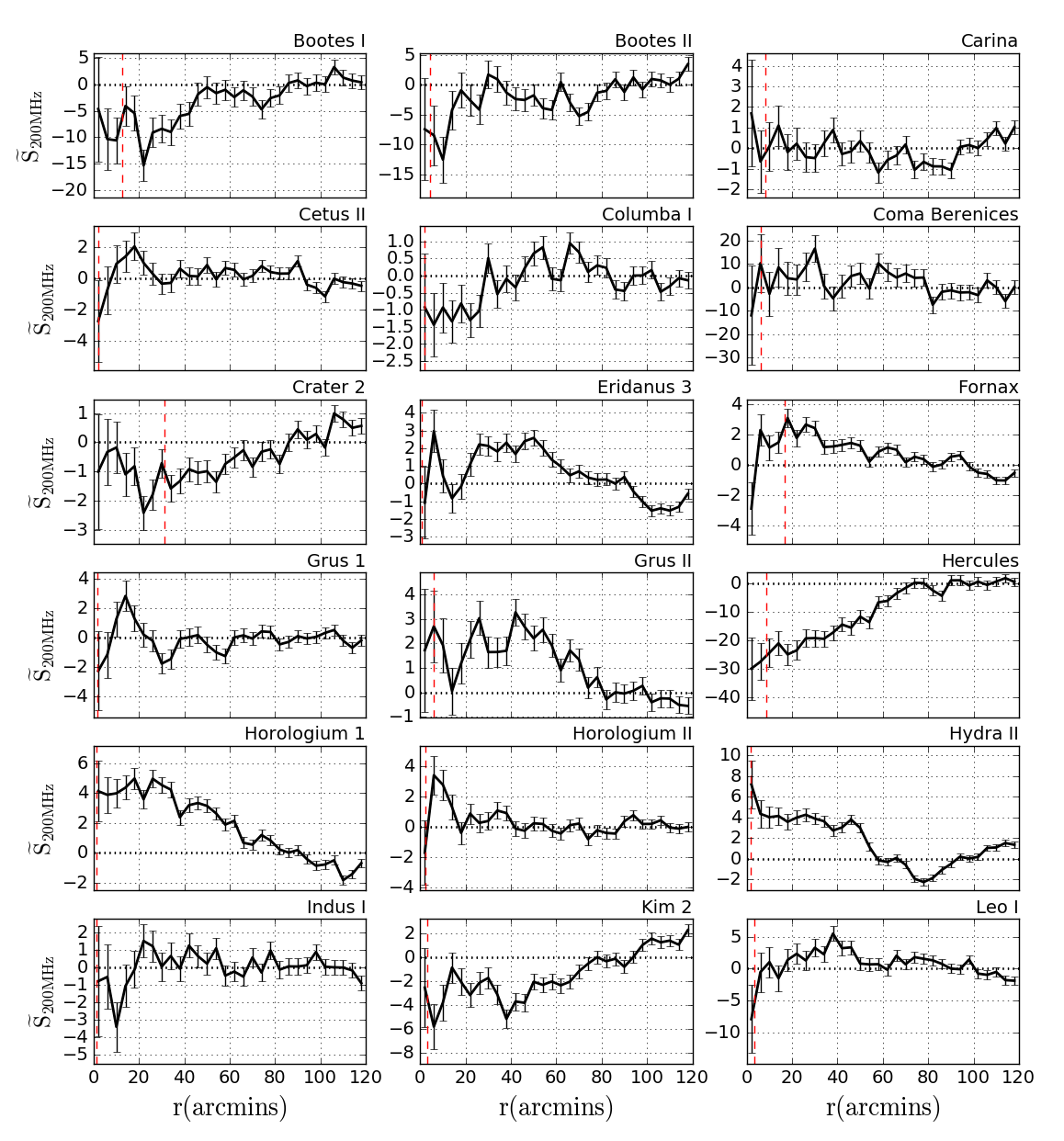}
    \caption{Flux density as a function of radius at 170\,--\,231\,MHz stacked frequency band for all 33 dSph galaxies in the sample. The flux density measurements here are the median flux density in units \mJybeam within a given annulus bin. The error bars are the average RMS noise within the annulus. Each radial annulus bin has a width of 4$'$. The profiles have been normalised by subtracting away the mean background level from $60' < r < 120'$. The red dashed line in each plot indicates the half-light radius (effective optical radius) for each galaxy \citep{McConnachie2012,Torrealba2016}.}
    \label{fig:galaxy_profiles}
\end{figure*}

Figure \ref{fig:galaxy_profiles} shows the intensity profiles for the 33 dSph galaxies in the sample. The profiles have been normalised to the respective background levels of their fields.

\begin{figure*}
    \centering
    \includegraphics[width=0.9\textwidth]{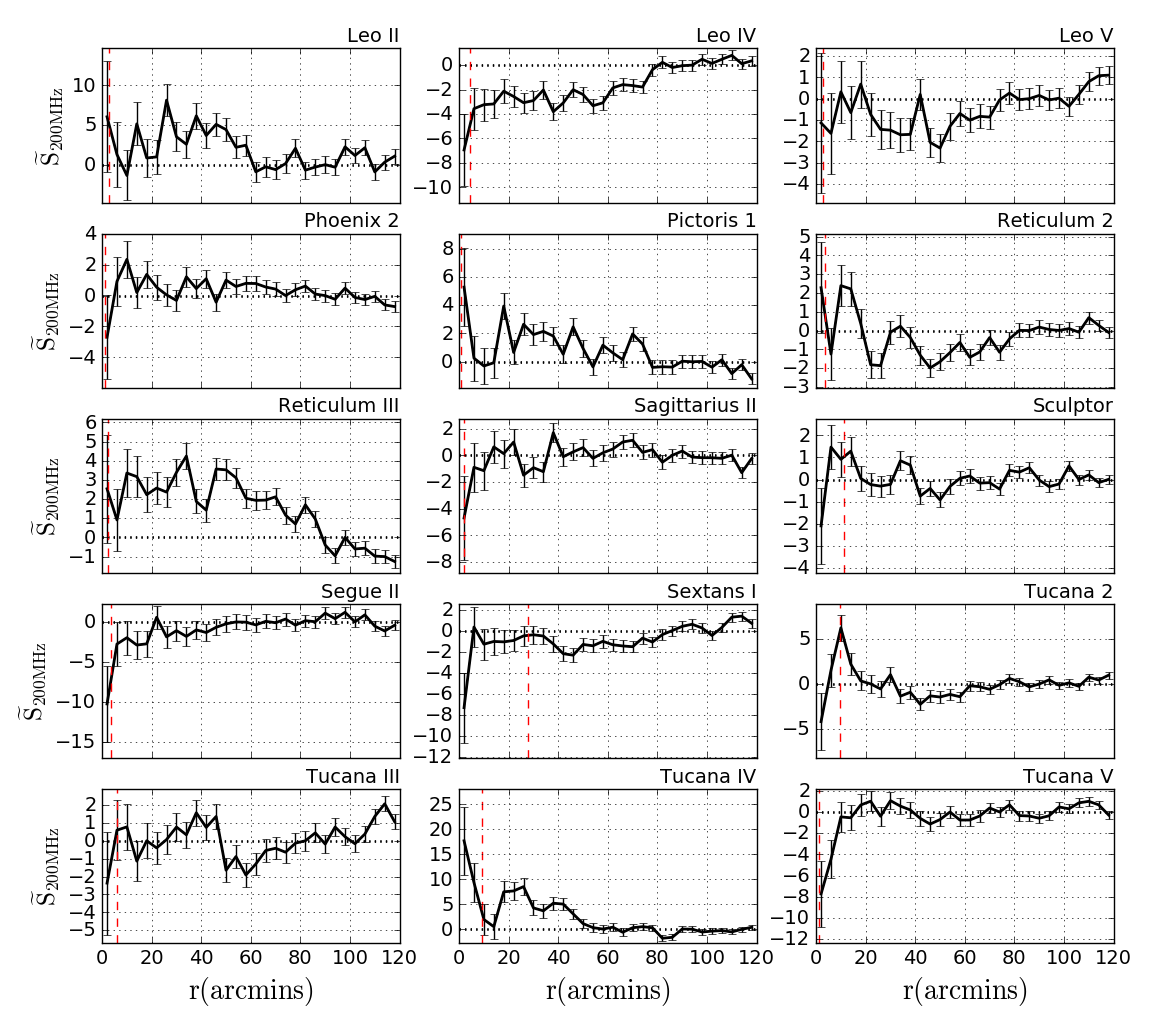}
    \caption{(continued).}
\end{figure*}

\bsp	
\label{lastpage}
\end{document}